\documentclass[showpacs,pra,aps,superscriptaddress,floatfix,tightenlines,twocolumn]{revtex4}

\usepackage{amsmath}
\usepackage{amssymb}
\usepackage{amsfonts}
\usepackage{hyperref}
\usepackage{graphicx}
\usepackage[dvips]{color}
\usepackage{bm}

\begin{document}
\title{Fitting magnetic field gradient with Heisenberg-scaling accuracy}
\author{Yong-Liang Zhang}
\affiliation{School of Physics, Peking University, Beijing 100871, China}
\author{Huan Wang}
\affiliation{School of Civil Engineering and Mechanics, Lanzhou University, Lanzhou 730000, China}
\author{Li Jing}
\affiliation{School of Physics, Peking University, Beijing 100871, China}
\author{Liang-Zhu Mu\footnote{muliangzhu@pku.edu.cn}}
\affiliation{School of Physics, Peking University, Beijing 100871, China}
\author{Heng Fan\footnote{hfan@iphy.ac.cn}}
\affiliation{Institute of Physics, Chinese Academy of Sciences, Beijing 100190, China}
\affiliation{Collaborative Innovation Center of Quantum Matter, Beijing, China}
\date{\today}

\begin{abstract}
We propose a quantum fitting scheme to estimate the magnetic field gradient with $N$-atom spins preparing in W state,
which attains the Heisenberg-scaling accuracy.
Our scheme combines the quantum multi-parameter estimation and the least square linear fitting method to
achieve the quantum Cram\'{e}r-Rao bound (QCRB). We show that the estimated quantity achieves the Heisenberg-scaling accuracy.
In single parameter estimation with assumption that the magnetic field is strictly linear,
two optimal measurements can achieve the identical Heisenberg-scaling accuracy.
Proper interpretation of the super-Heisenberg-scaling accuracy is presented.
The scheme of quantum metrology combined with data fitting provides a new method
in fast high precision measurements.
\end{abstract}
\maketitle

\emph{Introduction.}---Magnetometry is important for mineral exploration and probing moving magnetic objects.
High precision magnetometry \cite{budker,Vengalattore,shah,wasilewski,wolfgram,horrom,sewell} also has wide
applications in modern sciences and technologies, such as in nuclear magnetic resonance (NMR) \cite{greenberg}, magnetic resonance imaging (MRI) \cite{lauterbur,mansfield}, biomedical science \cite{matti} and quantum control \cite{grinolds}.
In general, the quantity interested is not the absolute strength of magnetic field but its difference and gradient. A standard measuring instrument for determining the gradient is differential atom interferometry, which utilizes two completely polarized atomic ensembles. Recently, quantum-enhanced measurements of magnetic field gradient have been proposed \cite{Eckert,htng,lanz,kim}.

It is by now well established that quantum metrology has advantages in enhancing precision of estimation \cite{Giovannetti3}
which is beyond the classical method.
In quantum metrology, the general framework for precision bound of estimation has been proposed and developed in Refs.\cite{fisher,cramer,helstrom, holevo,caves,gill,Escher}, which is based on quantum Fisher information and Cram\'{e}r-Rao inequality. The precision of estimation depends on the amount of resources employed in the scheme, which might be for instance the number $N$ of identical probes or the energy of probing field. The standard quantum limit,
a consequence of the central limit theorem for statistics, shows that the precision is proportional to $1/\sqrt{N}$. With quantum strategies such as entanglement
and squeezing applied, one may attain better accuracy scaling as $1/N$, which is the ultimate limit of precision named as Heisenberg limit. The NOON and GHZ states have been demonstrated to be able to provide a Heisenberg-limit sensitivity in some schemes \cite{bollinger,huelga,dowling,Giovannetti1,Giovannetti2}.
Also some experiments have implemented the quantum enhanced metrology \cite{Nagata,higgins,Kacprowicz,afek,xiang,Yonezawa}.

In this work, we propose a quantum scheme of multi-parameter estimation
to detect the gradient of magnetic field by employing $N$-atom spins.
These atoms are initially prepared in W state, a genuine multipartite entangled state that can be generated in spin chain \cite{Burb} and has been experimentally produced by trapped ions \cite{haffner}. These technologies can be utilized to implement our scheme in experiment.
By applying the quantum enhanced multi-parameter estimation to the least square linear fitting (LSLF) method,
we show that our scheme saturates the QCRB with Heisenberg-scaling accuracy. Let us highlight
some advantages of this scheme: (i) Our scheme does not depend on the prior assumed
linear assumption for the magnetic field, we essentially apply the reliable LSLF method.
(ii) We use the quantum multi-parameter estimation scheme which is robust
to accidental random errors in few parameters. Also this simultaneous estimation
scheme is in principle faster than repeated individual estimations.
(iii) This is a general quantum fitting method and can be applied to measure other physical
quantities with various fitting functions.
We also discuss that even if the linearity of the magnetic field is prior assumed,
the Heisenberg limit is always satisfied.

\emph{Local estimation theory.}---
We first present a brief review of local estimation theory, the Fisher information
and Cram\'{e}r-Rao inequality \cite{fisher,cramer,helstrom, holevo,caves,gill,Escher}.

Considering a curve $\hat\rho (\mathbf{y})$ characterizing dynamical process on the space of density matrix, the problem of determining the value of the parameter vector $\mathbf{y}=(y_1, y_2, \cdots, y_N)^T $ is a fundamental problem of statistical inference based on the experimental results. Before the measurements, we know that an observable random variable $\xi$ carries information about the unknown parameter vector $\mathbf{y}$, which is described by the smooth probability distribution $p(\xi|\mathbf{y})$. The normalization is $ \int d\xi  p(\xi|\mathbf{y}) =1$, and $\xi$ could be discrete or multivariate although it is written here as a single continuous real variable.

\begin{figure}[t]
  \includegraphics[width=0.45\textwidth]{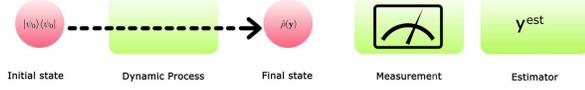}\\
  \caption{Scheme of quantum parameter estimation. The finial state $\hat\rho(\mathbf{y})$,
  evolved from a known initial state allowed
  by quantum mechanics, carries about the parameter vector characterizing dynamical process, and $\mathbf{y}^{est}$ is
  obtained from the measurement results performed on the final state.}\label{1}
\end{figure}

Then we take a random sample of size $\nu$ to estimate the parameter vector $\mathbf{y}$ via comparing the ratio of observed measurement outcomes with the probability distribution. An essential premise of effective deterministic estimation is requiring that the smooth map $p(\xi|\mathbf{y})\leftrightarrow \mathbf{y}$ is bijective. In order to avoid the periodical problems of determining the parameters $y_i$, it is generally assumed that all components $y_i$ are small, which is called local estimation. For an effective deterministic observable random variable $\xi$, one estimates the parameter vector $\mathbf{y}$ via funtions $y^{est}_i=y^{est}_i(\xi_1,\xi_2,\cdots, \xi_\nu)$ based on experimental results. The general framework of quantum parameter estimation is shown in FIG. \ref{1}. Then the expectation and covariance matrix of estimation are
\begin{align}
&\langle y^{est}_i\rangle = \int d\xi_1\cdots d\xi_\nu p(\xi_1|\mathbf{y})\cdots p(\xi_\nu|\mathbf{y}) y^{est}_i, \label{expectation} \\
&[\mathbf{Cov}(\mathbf{y}^{est})]_{m,n}=\int d\xi_1\cdots d\xi_\nu p(\xi_1|\mathbf{y})\cdots p(\xi_\nu|\mathbf{y})\nonumber \\
&\times( y^{est}_m -\langle  y^{est}_m \rangle) ( y^{est}_n -\langle  y^{est}_n \rangle).
\end{align}

Taking the partial derivative of Eq.(\ref{expectation}) with respect to $y_j$ and combining them into a bilinear quadratic form via two  arbitrary real vectors $\boldsymbol{\alpha}=(\alpha_1, \alpha_2, \cdots, \alpha_N)^T, \boldsymbol{\beta}=(\beta_1, \beta_2, \cdots, \beta_N)^T $, we obtain
\begin{align}\label{cov}
&\int d\xi_1\cdots d\xi_\nu p(\xi_1|\mathbf{y})\cdots p(\xi_\nu|\mathbf{y})\Big( \sum_{j=1}^N \alpha_j(\sum_{k=1}^\nu \frac{ \partial \ln p(\xi_k|\mathbf{y}) }{\partial y_j})\Big) \nonumber\\
&\times \Big(\sum_{i=1}^N \beta_i( y^{est}_i -\langle  y^{est}_i \rangle)\Big)= \sum_{i,j=1}^N \alpha_j\frac{ \partial \langle  y^{est}_i \rangle }{\partial y_j}\beta_i.
\end{align}
Applying the Cauchy-Schwarz inequality to Eq.(\ref{cov}) yields Cram\'{e}r-Rao bound \cite{fisher,cramer,helstrom, holevo,caves,gill}
\begin{align}\label{bound1}
\nu(\boldsymbol{\alpha}^T\boldsymbol{\mathcal{F}}(\mathbf{y})\boldsymbol{\alpha} ) (\boldsymbol{\beta}^T \mathbf{Cov}(\mathbf{y}^{est}) \boldsymbol{\beta})\geq \Big(\sum_{i,j=1}^N \alpha_j\frac{ \partial \langle  y^{est}_i \rangle }{\partial y_j}\beta_i\Big)^2,
\end{align}
where the Fisher information matrix (FI) is defined by
\begin{align} \label{fishermatrix}
[\boldsymbol{\mathcal{F}}(\mathbf{y})]_{m,n}=\int d\xi p(\xi|\mathbf{y}) \frac{ \partial \ln p(\xi|\mathbf{y}) }{\partial y_m}\frac{ \partial \ln p(\xi|\mathbf{y}) }{\partial y_n}.
\end{align}
Based on Eq.(\ref{bound1}), for all $\boldsymbol{\alpha}$, there exits $\boldsymbol{\beta}$ s.t. $(\boldsymbol{\alpha}^T\boldsymbol{\mathcal{F}}(\mathbf{y})\boldsymbol{\alpha})(\boldsymbol{\beta}^T \mathbf{Cov}(\mathbf{y}^{est}) \boldsymbol{\beta})>0$, and because $\boldsymbol{\beta}^T \mathbf{Cov}(\mathbf{y}^{est}) \boldsymbol{\beta}\geq 0$, then we find that the Fisher information matrix $\boldsymbol{\mathcal{F}}(\mathbf{y})$ is positive. Noticing that Eq.(\ref{bound1}) only holds for effective deterministic estimation, the Fisher information matrix defined by Eq.(\ref{fishermatrix}) is merely positive semi-definite for arbitrary observable random variables.

The asymptotic theory of maximum-likelihood estimation states that \cite{fisher,caves,gill}, in the approximate sense for large $\nu$, the estimation achieves the Cram\'{e}r-Rao bound and is unbiased locally, i.e. $\langle  y^{est}_i \rangle=y_i$, where $\mathbf{Cov}(\mathbf{y}^{est})$ is the matrix describing the deviation between the estimated values and real values. Thus for unbiased effective deterministic estimation, the Cram\'{e}r-Rao inequality can be written as \cite{helstrom,gill}
\begin{align}
\mathbf{Cov}(\mathbf{y})-[\nu\boldsymbol{\mathcal{F}}(\mathbf{y})]^{-1}\geq 0,
\end{align}
which means that it is a positive semi-definite matrix.

For quantum mechanics, the generalized measurement performed on the density matrix $\hat{\rho}(\mathbf{y})$ is described by a set of of non-negative Hermitian operators $ \hat{E}(\xi)$ \cite{nielsen}, which are complete in the sense that $\int d\xi \hat E(\xi)=\hat{\mathbb{I}}=\text{(unit operator)}$. And the probability distribution for measurement outcomes $\xi$ is given by $p(\xi|\mathbf{y})=Tr[ \hat E (\xi)\hat{\rho}(\mathbf{y})]$. As proven in \cite{caves}, we have
\begin{align}\label{quantumfisher}
\boldsymbol{\alpha}^T\boldsymbol{\mathcal{F}}(\mathbf{y})\boldsymbol{\alpha}\leq \boldsymbol{\alpha}^T\boldsymbol{\mathcal{F}}_Q(\mathbf{y})\boldsymbol{\alpha}, \forall\boldsymbol{\alpha},
\end{align}
where $\boldsymbol{\mathcal{F}}_Q(\mathbf{y})$ is the so-called quantum Fisher information (QFI) matrix defined as \cite{helstrom, holevo,gill}
\begin{align}
[\boldsymbol{\mathcal{F}}_Q(\mathbf{y})]_{m,n}=Tr[\hat{\rho}(\mathbf{y})\frac{\hat{L}_m\hat{L}_n+\hat{L}_n\hat{L}_m}{2}],
\end{align}
where these Hermitian operators are the so-called symmetric logarithmic derivatives, defined by the following equation
\begin{align}
\frac{\partial\hat{\rho}(\mathbf{y})}{\partial y_m}=\frac{\hat{L}_m\hat{\rho}(\mathbf{y})+\hat{\rho}(\mathbf{y}))\hat{L}_m}{2}.
\end{align}
The sufficient and necessary conditions for equality holding in Eq.(\ref{quantumfisher}) are
\begin{align}
\hat{E}(\xi)^{1/2}\Big(\sum_{m=1}^N \alpha_m \hat{L}_m -\lambda(\xi,\boldsymbol{\alpha}) \mathbb{I}\Big)\hat{\rho}(\mathbf{y})^{1/2}=0, \forall \xi, \forall\boldsymbol{\alpha},
\end{align}
where $\lambda(\xi,\boldsymbol{\alpha})=Tr[\hat{\rho}(\mathbf{y})\hat E (\xi)\sum_{m=1}^N \alpha_m \hat{L}_m ]/Tr[\hat E (\xi)\hat{\rho}(\mathbf{y})]$ is real. For single parameter estimation, the equality in Eq.(\ref{quantumfisher}) can always be satisfied by choosing the Hermitian operators to be one-dimensional projectors onto a complete set of orthonormal eigenstates of $\hat L$ \cite{caves}. Thus quantum Fisher information is the maximum of Fisher information over all possible measurement strategies \cite{caves,Escher}, i.e. $\mathcal{F}_Q=\max_{ \{\hat{E}(\xi) \} }\mathcal{F}$. For multi-parameter estimation, the equality in Eq.(\ref{quantumfisher}) generally is not achievable, which means that the quantum Cram\'{e}r-Rao inequality $\mathbf{Cov}(\mathbf{y})-[\nu\boldsymbol{\mathcal{F}}_Q(\mathbf{y})]^{-1}\geq 0$ cannot always be saturated \cite{helstrom,holevo,gill,vaneph,genoni,humphreys,yue}.

\begin{figure}[t]
  \includegraphics[width=0.45\textwidth]{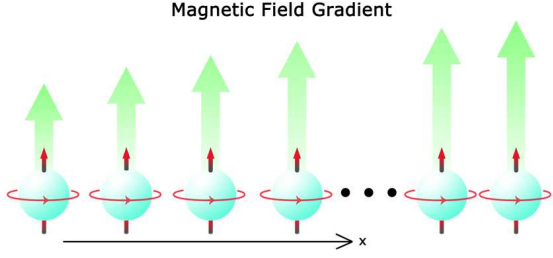}\\
  \caption{The atomic spin chain is coupled to a magnetic field, where each atom is separated with a distance $a$ in the x-direction.}\label{2}
\end{figure}

\emph{Multi-parameter estimation combined with the least square linear fitting method.}---Now, we consider the
problem of measuring the gradient of a magnetic field.
Our scheme is to estimate the strength of magnetic field at different locations through quantum measurements
and then to apply the LSLF method. We employ a $N$-atom spin chain as the probe, as shown in FIG. \ref{2}, to estimate the magnetic field gradient, where the $j$-th atom is located at $x_j=x_1+(j-1)a, (j=1,2,\cdots, N)$ and the uncertainty of the location $x_j$ can be neglected.
The Hamiltonian describes that each atom with two hyperfine spin states is coupled to the local magnetic field,
and it takes the form,
\begin{align}
\hat{H}=-\hbar \sum_{j=1}^N \gamma B_j \hat{\sigma}_z^j,
\end{align}
where $B_j$ and $\hat{\sigma}_z^j$ are the magnetic field and Pauli operator of atom $j$, and each atom has the same gyromagnetic ratio $\gamma$. The task of our scheme is to obtain optimal uncertainty bound of estimating the magnetic field gradient $G$ that quantum mechanics permitted.
Initially, the atomic spins are prepared in a W state $|\psi_0\rangle=\frac{1}{\sqrt{N}}\sum_{j=1}^N |w_j\rangle$
by symmetric consideration, where $|w_j\rangle=|1\rangle_j\prod_{j'\neq j} |0\rangle_{j'}$. For this closed quantum system, then the quantum state evolves under the action of magnetic field as $\hat{\rho}(\mathbf{B})=U(\mathbf{B})|\psi_0\rangle \langle \psi_0| U^\dag(\mathbf{B})$, where $U(\mathbf{B})=e^{-i\hat{H}t/\hbar}$ due to
Sch\"odinger equaiton. The initial pure state acquired multiple phases through the unitary transformation is given by
\begin{align}
|\psi(t,\mathbf{B})\rangle=\frac{1}{\sqrt{N}}\sum_{j=1}^N e^{-i2\gamma t B_j} |w_j\rangle.
\end{align}
Because of an overall unobservable phase, it is proper to think that $B_1=0$ always holds and
the covariance matrix $\mathbf{Cov}(\mathbf{B})$ is size $(N-1)\times(N-1)$. Generalizing the expression of estimation for unitary dynamical processes \cite{Escher}, the QFI matrix is given by $[\boldsymbol{\mathcal{F}}_Q(\mathbf{B})]_{m,n}=2[\langle \hat{h}_m \hat{h}_n+\hat{h}_n \hat{h}_m\rangle_0-2\langle \hat{h}_m\rangle_0\langle \hat{h}_n\rangle_0$] \cite{yue}, where $\hat{h}_m=i\frac {\partial U^\dag(\mathbf{B})}{\partial B_m} U(\mathbf{B})$.
By straightforward calculations, the QFI matrix and its inverse associated with the estimation of the magnetic fields $B_j$ in our scheme is
\begin{align}
[\boldsymbol{\mathcal{F}}_Q(\mathbf{B})]_{m,n}&=\frac{16\gamma^2 t^2}{N^2}(N\delta_{m,n}-1),\\
[\boldsymbol{\mathcal{F}}_Q(\mathbf{B})]^{-1}_{m,n}&=\frac{N}{16\gamma^2 t^2}(\delta_{m,n}+1).
\end{align}
Applying the LSLF method, we have the fitting gradient of the magnetic field as,
\begin{align}
G=\frac{\sum_{i=1}^N (x_i-\overline{x})(B_i-\overline{B})}{\sum_{i=1}^N (x_i-\overline{x})^2}=\sum_{i=1}^N c_i B_i,
\end{align}
where $\overline{x}=(\sum_{i=1}^N  x_i)/N, \overline{B}=(\sum_{i=1}^N  B_i)/N$, and $c_i=\frac{6(2i-N-1)}{a(N-1)N(N+1)}$.
Since the uncertainties of $x_j$ are neglected, the quantum Cram\'{e}r-Rao inequality gives a lower bound on the variance of the magnetic field gradient
\begin{align}
\sigma_G\geq \sqrt{\sum_{m,n=2}^N c_m [\nu\boldsymbol{\mathcal{F}}_Q(\mathbf{B})]^{-1}_{m,n} c_n }=\frac{1}{2\gamma ta}\sqrt{\frac{3}{\nu(N^2-1)}}.
\end{align}
This bound clearly goes beyond the quantum standard limit and achieves Heisengberg-scaling accuracy for large $N$.

In this scheme, we construct two von Neumann measurements labeled by $a,b$ respectively, $\hat{E}^{a(b)}(\xi)=|\Pi^{a(b)}_\xi\rangle\langle \Pi^{a(b)}_\xi|$,
to be performed on the atomic spin chain as the following forms,
\begin{align}
|\Pi^a_0\rangle &=|\Pi^b_0\rangle= \frac{1}{\sqrt{N}}\sum_{j=1}^N |w_j\rangle, \\
|\Pi^a_k\rangle&=\frac{1}{\sqrt{N}}\sum_{j=1}^N e^{i\frac{2\pi k}{N}(j-1)}|w_j\rangle, \\
|\Pi^b_k\rangle&= \sqrt{\frac{k}{k+1}}\Big( \frac{1}{k}\sum_{j=1}^k |w_j\rangle-|w_{k+1}\rangle \Big),
\end{align}
where $k=1,2,\cdots, N-1$. Both of these two sets of quantum states are orthonormal eigenstates of the coherence operator expressed as $\hat{C}=(N-1)\hat{E}^{a(b)}(0)-\sum_{\xi=1}^{(N-1)}\hat{E}^{a(b)}(\xi)$, see Ref.\cite{kim}.
We can obtain the Fisher information matrices of these two sets of measurements, respectively,
\begin{align}
\lim_{\{B_j\rightarrow (j-1)Ga\}}[\boldsymbol{\mathcal{F}}^a(\mathbf{B})]_{m,n}&=\frac{8\gamma^2 t^2}{N}(\delta_{m,n}-\delta_{m+n,N+1}),\\
\lim_{\mathbf{B}\rightarrow\mathbf{0}}[\boldsymbol{\mathcal{F}}^b(\mathbf{B})]_{m,n}
&=\frac{16\gamma^2 t^2}{N^2}(N\delta_{m,n}-1),
\end{align}
see supplementary material for detailed calculations \cite{supple}.

For the first set of measurements, the Fisher information matrix is merely positive semi-definite and irreversible, which confirms that it is not an effective deterministic estimation. Applying Fourier transformation, we have $\lambda_\xi=\frac{1}{\sqrt{N}}\sum_{j=1}^N e^{-i2\gamma tB_j-\frac{i2\pi\xi}{N}(j-1)}$ and $e^{-i2\gamma tB_j}=\frac{1}{\sqrt{N}}\sum_{\xi=1}^N e^{\frac{i2\pi\xi}{N}(j-1)}\lambda_\xi$. Because $p(\xi|\mathbf{B})=|\lambda_\xi|^2/N $, it is impossible to estimate $\lambda_\xi$ and the magnetic field $B_j$ from the probability distribution associated with experimental outcomes. For the second set of measurements, which yields the QFI matrix, the probability of each outcome is transparently related to the magnetic field $\mathbf{B}$, with $p(1|\mathbf{B})$ involving only $B_2$, $p(2|\mathbf{B})$ involving only $B_2,B_3$, and so on \cite{humphreys}. This suggests that the estimator could effectively determine the magnetic field $\mathbf{B}$. Based on the results of asymptotically large $\nu$ independent experiments, this set of measurements is optimal which can locally achieve the Heisenberg-scaling quantum Cram\'{e}r-Rao bound.

\emph{Single parameter estimation with linear assumption.}---
If we assume that the magnetic field satisfies the linear condition $B_j=B_1+G(j-1)a$,
the single parameter representing gradient $G$ of magnetic field needs be estimated.
In this case, the unitary transformation for the atomic spin chain is $\hat{U}(G)=e^{-i\hat{H}t/\hbar}$,
and the QFI can be expressed as \cite{Escher}
\begin{align}\label{quantum fisher}
\mathcal{F}_Q=4[\langle\psi_0|\hat{h}(G)^2|\psi_0\rangle-\langle\psi_0|\hat{h}(G)|\psi_0\rangle^2],
\end{align}
where $\hat{h}(G)=i\frac{d \hat{U}^\dag(G)}{dG}\hat{U}(G)=\gamma ta \sum_{j=1}^N (j-1)\hat{\sigma}_z^j $.
Applying this equation, we obtain $\mathcal{F}_Q=\frac{(2\gamma ta)^2}{3}(N^2-1)$.

We show in supplementary material \cite{supple} that the Fisher information of previously proposed two sets of measurements
are identical,
\begin{align}
\mathcal{F}^a(G)&=\frac{(2\gamma ta)^2}{3}(N^2-1),\\
\mathcal{F}^b(G)&\overset{\gamma taG\ll 1}{\approx}\frac{(2\gamma ta)^2}{3}(N^2-1).
\end{align}
These two sets measurements are optimal because they both yield the QFI.
It is straightforward to determine that the quantum Cram\'{e}r-Rao bound $\sigma^{a(b)}_{G}=\frac{1}{2\gamma ta}\sqrt{\frac{3}{\nu(N^2-1)}}$
which is exactly the same as the Heisenberg-scaling accuracy for scheme of the multi-parameter estimation.
For measurements $\hat{E}^{a}(\xi)=|\Pi^{a}_\xi\rangle\langle \Pi^{a}_\xi|$, the probability distribution $p^a(\xi|G)$ is clearly peaked around $(-\xi/N+ j)\pi/(\gamma t a) $ with approximate width $\pi/(N\gamma t a)$, where $j$ is an arbitrary integer.
If the condition $0<G<\pi/(\gamma t a)$ is satisfied, one can successfully estimate $G$ with Heisenberg-scaling accuracy.
This measurement is essentially a quantum Fourier algorithm for phase estimation \cite{nielsen,chuang}.  For measurements $\hat{E}^{b}(\xi)=|\Pi^{b}_\xi\rangle\langle \Pi^{b}_\xi|$, the Heisenberg-scaling accuracy can only be reached locally, i.e., the unknown parameter satisfies $\gamma taG\ll 1$.

\emph{The trick of super-Heisenberg-scaling accuracy.}---
With the prior assumed linear condition for magnetic gradient,
we can find an optimal initial pure state which maximizes the QFI.
Considering the initial state as $|\psi_0\rangle=\sum_{i_1,i_2,\cdots,i_N=0}^1 x_{i_1i_2\cdots i_N}|i_1i_2\cdots i_N\rangle$,
we then find,
\begin{align}
\mathcal{F}_Q=(2\gamma ta)^2\Big[ \sum_{i_1,i_2,\cdots,i_N=0}^1 |x_{i_1i_2\cdots i_N}|^2 a_{i_1i_2\cdots i_N}^2-\nonumber \\
\Big(\sum_{i_1,i_2,\cdots,i_N=0}^1 |x_{i_1i_2\cdots i_N}|^2 a_{i_1i_2\cdots i_N}\Big)^2\Big],
\end{align}
where $a_{i_1i_2\cdots i_N}=\sum_{j=1}^N (j-1)(-1)^{i_j}$. Using the mathematical proposition in \cite{supple}, the GHZ state
$|GHZ\rangle=\frac{1}{\sqrt{2}}(|00\cdots 0\rangle+|11\cdots 1\rangle)$ has the maximum QFI $\mathcal{F}_Q|_{GHZ}=N^2(N-1)^2(\gamma ta)^2$.
For the situation that $N$ is even, the QFI of the NOON state $|NOON\rangle=\frac{1}{\sqrt{2}}(|\underbrace{00\cdots 0}_{N/2}\underbrace{11\cdots 1}_{N/2} \rangle +|\underbrace{11\cdots 1}_{N/2}\underbrace{00\cdots 0}_{N/2}\rangle)$ is $\mathcal{F}_Q|_{NOON}=N^4(\gamma ta)^2/4$. The quantum Cram\'{e}r-Rao bound $\sigma_{G}=\frac{1}{\sqrt{\nu\mathcal{F}_Q}}$ for GHZ state and NOON state are both super-Heisenberg-scaling accuracy for large $N$, and these bounds can be achievable via parity measurement \cite{bollinger,dowling,ylzhang,kessler,komar}. Taking account of the periodical problem of estimating an unknown phase $0<\phi<\pi$ in quantum enhanced metrology employing GHZ state or NOON state, one might yield the estimation with Heisenberg-limited accuracy up to a logarithmic correction \cite{burgh,kessler,komar}.

However, close scrutiny of this $\frac{1}{N^2}$-scaling accuracy reveals that it is completely a trick for detecting the gradient which is based on strictly linear hypothesis of magnetic field. For GHZ state, $|\psi(t,\mathbf{B})\rangle=\frac{1}{\sqrt{2}}(|00\cdots 0\rangle+e^{-i2\gamma t N \overline{B}}|11\cdots 1\rangle),$ where $\overline{B}=(\sum_{j=1}^N B_j)/N$, and $\sigma_{\overline{B}}=\frac{1}{2\gamma t \sqrt{\nu N^2}}$ is a Heisenberg limit.
For NOON state, $|\psi(t,\mathbf{B})\rangle=\frac{1}{\sqrt{2}}(|00\cdots 011\cdots 1\rangle +e^{-i\gamma t N \Delta \overline{B}}|11\cdots 100\cdots 0\rangle)$, where $\Delta \overline{B}=(B_{\frac{N}{2}+1}+\cdots+B_N-B_1-\cdots B_{\frac{N}{2}})/(N/2)$, and $\sigma_{\Delta \overline{B}}=\frac{1}{\gamma t \sqrt{\nu N^2}}$,
still the Heisenberg limit is obtained. Thus we conclude that the Heisenberg limit is still true with proper interpretation.

\emph{Conclusions.}---
Determining the gradient of magnetic field is inherently a multi-parameter estimation problem.
We employ quantum enhanced multi-parameter estimation and the least square linear fitting method
to achieve the Heisenberg-scaling accuracy. Our scheme provides attainable high precision in magnetometry.
This proposal is the first data fitting scheme possessing Heisenberg-scaling accuracy.
This opens a new avenue for the investigations of general data fitting problems.

This work was supported by the ``973'' Program
(2010CB922904), NSFC (11175248), NFFTBS (J1030310, J1103205),
grants from the Chinese Academy of Sciences, and the Chun-Tsung
scholar fund of Peking University.

\begin{widetext}
\section{Supplementary Material}

\subsection{Calculation of Fisher information matrixes in multi-parameter estimation}
For von Neumann measurements $|\Pi_\xi\rangle=\sum_{j=1}^N U_{(\xi+1),j}|w_j\rangle$, the probability distribution of each measurement result is
\begin{align}
p(\xi|\mathbf{B})=Tr[\hat E(\xi) \hat{\rho}(\mathbf{B})]=\sum_{\mu,\nu=1}^N z_\mu \widetilde{z}_\nu U_{(\xi+1),\mu} \widetilde{U}_{(\xi+1),\nu},
\end{align}
where $z_\mu=\frac{1}{\sqrt{N}}e^{i2\gamma t B_\mu}$, $\widetilde{z}$ denote the complex conjugate of $z$ and U is a unitary matrix, i.e., $\sum_{k=1}^N U_{k,\mu}\widetilde{U}_{k,\nu}=\delta_{\mu,\nu}$. Then the Fisher information matrix is
\begin{align}
[\boldsymbol{\mathcal{F}}(\mathbf{B})]_{m,n}&=\sum_{\xi=0}^{N-1} \frac{ \partial_{B_m} p(\xi|\mathbf{B})\partial_{B_n} p(\xi|\mathbf{B})}{p(\xi|\mathbf{B})}\\
&=4\gamma^2 t^2 \sum_{k=0}^{N} \frac{\sum_{\mu,\nu=1}^{N}[-iz_\mu \widetilde{z}_m U_{k,\mu} \widetilde{U}_{k,m}+i \widetilde{z}_\mu z_m \widetilde{U}_{k,\mu}U_{k,m} ]\times [-iz_\nu \widetilde{z}_n U_{k,\nu} \widetilde{U}_{k,n}+i \widetilde{z}_\nu z_n \widetilde{U}_{k,\nu}U_{k,n} ] }{\sum_{\mu,\nu=1}^N z_\mu \widetilde{z}_\nu U_{k,\mu} \widetilde{U}_{k,\nu}} \\
&=4\gamma^2 t^2 \sum_{k=1}^{N} 2Re\Big[ \widetilde{z}_m z_n \widetilde{U}_{k,m} U_{k,n}-\widetilde{z}_m\widetilde{z}_n \widetilde{U}_{k,m} \widetilde{U}_{k,n} \frac{\sum_{\mu,\nu=1}^N z_\mu z_\nu U_{k,\mu} U_{k,\nu}}{\sum_{\mu,\nu=1}^N z_\mu \widetilde{z}_\nu U_{k,\mu} \widetilde{U}_{k,\nu}} \Big] \\
&=8\gamma^2 t^2 \Big[ \frac{1}{N}\delta_{m,n}-Re\Big(\widetilde{z}_m\widetilde{z}_n \sum_{k=1}^N\widetilde{U}_{k,m} \widetilde{U}_{k,n} \frac{\sum_{\mu,\nu=1}^N z_\mu z_\nu U_{k,\mu} U_{k,\nu}}{\sum_{\mu,\nu=1}^N z_\mu \widetilde{z}_\nu U_{k,\mu} \widetilde{U}_{k,\nu}} \Big) \Big]. \\
\end{align}

For measurements $\hat{E}^{a}(\xi)=|\Pi^{a}_\xi\rangle\langle \Pi^{a}_\xi|$, the unitary matrix is $U^a_{\mu,\nu}=\frac{1}{\sqrt{N}}e^{\frac{i2\pi(\mu-1)(\nu-1)}{N}}$. If one supposes that $B_j=(j-1)aG$, then
\begin{align}
\sum_{\mu=1}^{N} z_\mu U_{k,\mu} &=\frac{\sin N(\gamma t a G +\frac{(k-1)\pi}{N})}{N\sin(\gamma t a G +\frac{(k-1)\pi}{N})}e^{i(N-1)(\gamma t a G +\frac{(k-1)\pi}{N})},\\
\lim_{\{B_j\rightarrow (j-1)Ga\}}[\boldsymbol{\mathcal{F}}^a(\mathbf{B})]_{m,n}
&=\frac{8\gamma^2 t^2}{N}  \Big[ \delta_{m,n}-\frac{1}{N} Re\Big(e^{-i2\gamma t aG(m+n-N-1)}\sum_{k=1}^{N}e^{-\frac{i 2\xi(k-1)}{N}(m+n-N-1)} \Big) \Big] \\
&=\frac{8\gamma^2 t^2}{N}(\delta_{m,n}-\delta_{m+n,N+1})
\end{align}

For measurements $\hat{E}^{b}(\xi)=|\Pi^{b}_\xi\rangle\langle \Pi^{b}_\xi|$, the unitary matrix
degrades into an orthogonal matrix $O^b_{\mu,\nu}$, we have
\begin{align}
&\lim_{\mathbf{B}\rightarrow\mathbf{0}}\frac{\sum_{\mu,\nu=1}^N z_\mu z_\nu O^b_{k,\mu} O^b_{k,\nu}}{\sum_{\mu,\nu=1}^N z_\mu \widetilde{z}_\nu O^b_{k,\mu} O^b_{k,\nu}}=2\delta_{k,1}-1 ,\\
&\lim_{\mathbf{B}\rightarrow\mathbf{0}}[\boldsymbol{\mathcal{F}}^b(\mathbf{B})]_{m,n}
=\frac{16\gamma^2 t^2}{N^2}(N\delta_{m,n}-1).
\end{align}

\subsection{Calculation of Fisher information in single parameter estimation}
Since $\frac{\sin N(\gamma t a G +\frac{\xi\pi}{N})}{\sin(\gamma t a G +\frac{\xi\pi}{N})}=\sum_{m=-J}^J e^{i2m(\gamma ta+\frac{\xi\pi}{N})}$, where $J=(N-1)/2$, we calculate the probability distribution and the Fisher information of measurements $\hat{E}^{a}(\xi)=|\Pi^{a}_\xi\rangle\langle \Pi^{a}_\xi|$
\begin{align}
p^a(\xi|G)&=\frac{1}{N^2}|\sum_{j=1}^N e^{-i2(j-1)(\gamma taG+\frac{\xi \pi}{N})}|^2=\frac{1}{N^2}\frac{\sin^2N(\gamma t a G +\frac{\xi\pi}{N})}{\sin^2(\gamma t a G +\frac{\xi\pi}{N})},\\
\mathcal{F}^a (G)&=\sum_{\xi=0}^{N-1} \frac{1}{p^a(\xi|G)}[\frac{d p^a(\xi|G)}{d G}]^2=\frac{4}{N^2}\sum_{\xi=0}^{N-1} [\frac{d}{d G} \frac{\sin N(\gamma t a G +\frac{\xi\pi}{N})}{\sin(\gamma t a G +\frac{\xi\pi}{N})}]^2 \\
&=\frac{(2\gamma ta)^2}{N^2}\sum_{\xi=0}^{N-1} \sum_{m,m'=-J}^J 4mm' e^{i2(m-m')x+i2(m-m')\frac{\xi\pi}{N}}\nonumber \\
&=\frac{(2\gamma ta)^2}{N^2}\sum_{m,m'=-J}^J 4mm'  e^{i2(m-m')x}N\delta_{m,m'}=\frac{(2\gamma ta)^2}{3}(N^2-1).
\end{align}

The probability distribution and the Fisher information of measurements $\hat{E}^{b}(\xi)=|\Pi^{b}_\xi\rangle\langle \Pi^{b}_\xi|$ are
\begin{align}
&p^b(0|G)=\frac{1}{N^2}\frac{\sin^2(N\gamma t a G)}{\sin^2(\gamma t a G)} \overset{\gamma taG\ll 1}{\approx} 1 + \frac{(N^2-1)(\gamma taG)^2}{3},\\
&p^b(\xi|G)=\frac{\xi}{N(\xi+1)}\Big[1-2\cos
[(\xi+1) \gamma taG] \frac{\sin (\xi \gamma taG)}{\xi \sin( \gamma taG)} +\Big(\frac{\sin(\xi \gamma taG)}{\xi \sin( \gamma taG)}\Big)^2\Big]\overset{\gamma taG\ll 1}{\approx} \frac{\xi(1+\xi) (\gamma taG)^2}{N},\\
&\mathcal{F}^b (G)=\sum_{\xi=0}^{N-1} \frac{1}{p^b(\xi|G)}[\frac{d p^b(\xi|G)}{d G}]^2\overset{\gamma taG\ll 1}{\approx}
\frac{(2\gamma ta)^2}{3}(N^2-1)
\end{align}

\subsection{A mathematical proposition using Lagrange multiplier}
\begin{align}
\begin{cases}
f=\sum_j x_j^2a^2_j-(\sum_j x_j^2 a_j)^2\\
\phi=\sum_j x_j^2-1=0
\end{cases}
\end{align}
Using Lagrange multiplier, it is easy to prove that the funtion reaches the maximum $f_{max}=\frac{(a_M-a_m)^2}{4}$ when $x_M^2=x_m^2=1/2$, where $a_M=\max\{a_j\}$ and $a_m=\min\{a_j\}$.

\end{widetext}

\end{document}